\documentstyle[prl,aps]{revtex}
 
\begin{document}
 
\preprint{SLAC-PUB-8049, February 22, 1999 T/E}
 
\title{Inclusive Hadron Photoproduction from Longitudinally 
Polarized Protons and Deuterons}
 
\author{The E155 Collaboration \break
P.~L.~Anthony,$^{16}$  
R.~G.~Arnold,${^1}$
T.~Averett,$^{5,\diamond}$
H.~R.~Band,$^{21}$
M.~C.~Berisso,$^{12}$
H.~Borel,$^7$
P.~E.~Bosted,${^1}$
S.~L.~B${\ddot {\rm u}}$ltmann,$^{19}$
M.~Buenerd,$^{16,\dag}$
T.~Chupp,$^{13}$
S.~Churchwell,$^{12,\ddag}$
G.~R.~Court,$^{10}$
D.~Crabb,$^{19}$
D.~Day,$^{19}$
P.~Decowski,$^{15}$
P.~DePietro,$^1$
R.~Erbacher,$^{16,17}$
R.~Erickson,$^{16}$
A.~Feltham,$^{19}$
H.~Fonvieille,$^3$
E.~Frlez,$^{19}$
R.~Gearhart,$^{16}$
V.~Ghazikhanian,$^6$
J.~Gomez,$^{18}$
K.~A.~Griffioen,$^{20}$
C.~Harris,$^{19}$
M.~A. Houlden,$^{10}$
E.~W.~Hughes,$^5$
C.~E.~Hyde-Wright,$^{14}$
G.~Igo,$^6$
S.~Incerti,$^3$
J.~Jensen,$^5$
J.~R.~Johnson,$^{21}$
P.~M.~King,$^{20}$
Yu.~G.~Kolomensky,$^{5,12}$
S.~E.~Kuhn,$^{14}$
R.~Lindgren,$^{19}$
R.~M.~Lombard-Nelsen,$^7$
J.~Marroncle,$^7$
J.~McCarthy,$^{19}$
P.~McKee,$^{19}$
W.~Meyer,$^{4}$
G.~S.~Mitchell,$^{21}$
J.~Mitchell,$^{18}$
M.~Olson,$^{9,\Box}$
S.~Penttila,$^{11}$
G.~A.~Peterson,$^{12}$
G.~G.~Petratos,$^9$
R.~Pitthan,$^{16}$
D.~Pocanic,$^{19}$
R.~Prepost,$^{21}$
C.~Prescott,$^{16}$
L.~M.~Qin,$^{14}$
B.~A.~Raue,$^{8}$
D.~Reyna,$^{1,\flat}$
L.~S.~Rochester,$^{16}$
S.~Rock,$^1$
O.~A.~Rondon-Aramayo,$^{19}$
F.~Sabatie,$^7$
I.~Sick,$^2$
T.~Smith,$^{13}$
L.~Sorrell,$^1$
F.~Staley,$^7$
S.~St.Lorant,$^{16}$
L.~M.~Stuart,$^{16,\S}$
Z.~Szalata,$^1$
Y.~Terrien,$^7$
A.~Tobias,$^{19}$
L.~Todor,$^{14}$
T.~Toole,$^1$
S.~Trentalange,$^{6}$
D.~Walz,$^{16}$
R.~C.~Welsh,$^{13}$
F.~R.~Wesselmann,$^{14}$
T.~R.~Wright,$^{21}$
C.~C.~Young,$^{16}$
M.~Zeier,$^2$
H.~Zhu,$^{19}$
B.~Zihlmann,$^{19}$}
\address{
{$^{1}$American University, Washington, D.C. 20016}  \break
{$^{2}$Institut f${\ddot u}$r Physik der Universit${\ddot a}$t 
Basel, CH-4056 Basel, Switzerland} \break
{$^{3}$University Blaise Pascal, LPC IN2P3/CNRS F-63170 
Aubiere Cedex, France} \break
{$^{4}$Ruhr-Universit${\ddot a}$t Bochum, 
Universit${\ddot a}$tstr. 150, 
Bochum, Germany} \break
{$^{5}$California Institute of Technology, Pasadena, California 91125}
\break
{$^{6}$University of California, Los Angeles, California 90095}
\break
{$^{7}$DAPNIA-Service de Physique Nucleaire, CEA-Saclay,
F-91191 Gif/Yvette Cedex, France} \break
{$^{8}$Florida International University, Miami, Florida 33199.} \break
{$^{9}$Kent State University, Kent, Ohio 44242} \break
{$^{10}$University of Liverpool, Liverpool L69 3BX, United Kingdom } \break
{$^{11}$Los Alamos National Laboratory, Los Alamos, New Mexico 87545} \break
{$^{12}$University of Massachusetts, Amherst, Massachusetts 01003} \break
{$^{13}$University of Michigan, Ann Arbor, Michigan 48109} \break
{$^{14}$Old Dominion University, Norfolk, Virginia 23529} \break
{$^{15}$Smith College, Northampton, Massachusetts 01063} \break
{$^{16}$Stanford Linear Accelerator Center, Stanford, California 94309 } 
\break
{$^{17}$Stanford University, Stanford, California 94305} \break
{$^{18}$Thomas Jefferson National Accelerator Facility, Newport News, 
Virginia
23606} \break
{$^{19}$University of Virginia, Charlottesville, Virginia 22901} \break
{$^{20}$The College of William and Mary , Williamsburg, Virginia 23187} \break
{$^{21}$University of Wisconsin, Madison, Wisconsin 53706} \break
}
 
\maketitle
 
\begin{abstract}
We report measurements of the asymmetry $A_\|$ for inclusive hadron production
on  longitudinally polarized proton
and deuteron targets  by circularly polarized photons. The  
photons were produced via internal and external
bremsstrahlung from an electron beam of 48.35 GeV.
 Asymmetries for both positive and
negative signed hadrons, and a subset of identified pions, 
 were measured in the momentum 
range $10<P<30$ GeV at 2.75$^{\circ}$ and 5.5$^{\circ}$. 
Small non-zero asymmetries are observed for the proton, while the
deuteron results are consistent with zero. 
Recent calculations do not describe the data well.   
  \end{abstract}
 
 
%\narrowtext
\twocolumn
There has been much recent interest in the spin structure of the nucleon,
both theoretically and experimentally. The helicity-dependent parton 
distributions have been probed in a recent series of deep-inelastic 
polarized lepton-nucleon
scattering experiments at SLAC \cite{E155}, CERN \cite{SMC}, and 
DESY \cite{hermes}. These experiments are
primarily sensitive to the polarized quark densities, and the  sensitivity
to specific quark  flavors can be enhanced by detecting particular 
 mesons in coincidence with the scattered lepton (semi-inclusive
measurements). It has recently been suggested \cite{carlson,many}
 that extending such
measurements to $Q^2=0$ (the photoproduction limit) may reveal interesting
sensitivity to both the quark and gluon densities. This is because, in 
addition to contributions where a photon is absorbed on a quark,
photons can fuse with gluons in the nucleon to produce a 
quark-antiquark pair, and the analyzing power 
for this process is large. In the ideal 
experiment, a monochromatic circularly polarized photon would be absorbed
on a polarized proton or neutron, and an emitted pion or kaon would be
detected in the final state. 
 
In this Letter we report on measurements of inclusive hadron production
by a polarized bremsstrahlung photon beam impinging on polarized proton 
and deuteron targets. The measurements were taken concurrently with
inclusive electron scattering in SLAC E155 \cite{E155}. The polarized
electron beam of energy $E_0=48.35$ GeV and polarization $P_e=0.813\pm0.020$
passed through the polarized target, producing a bremsstrahlung photon beam 
with an effective  flux approximately given by $\Phi(k)=(t/2)(dk/k)$, 
where $k$ is the photon energy, and $t=0.04$ and 0.02 radiation
lengths (r.l.) for the NH$_3$ and LiD targets respectively. 
The electroproduction
of hadrons by electrons that scatter at close to zero degrees can be
considered to give an additional flux of approximately  $\Phi(k)=0.04(dk/k)$
in the effective radiator approximation.
 Since only
the helicity-dependent asymmetry for hadron photoproduction is measured
in this experiment, it is not important to know the magnitude of the photon
flux. The photon circular polarization 
is given by \cite{olson} $P_\gamma/P_e=y(4-y)/(4-4y+3y^2)$, where
$y=k/E_0$. This formula yields values of $P_\gamma/P_e$ of 1, 0.91, 0.64,
and 0.29 for $y=1$, 0.75, 0.5, and 0.25, respectively, illustrating that
the photons are the most polarized near the endpoint. 
 
The longitudinally polarized proton target was a 3-cm-long cell filled with
granules of $^{15}$NH$_3$ immersed in liquid He at  1 K in  a uniform 
magnetic field of 5 T.  The proton polarization varied from 0.6 to 0.9 during
the experiment, with a typical value of $P_t=0.8$. A small ($<2\%$) correction
was made to the asymmetry measurements for the polarization of the 
$^{15}$N nuclei. Crystals of $^6$LiD were used for the 
deuteron target, in which 
the $^6$Li nuclei were treated as an effective 
 polarized deuteron with 86\% of the
polarization of the free deuteron \cite{LiD}, which averaged 
about $P_t=0.22$ during the experiment. 
The polarization directions of both the NH$_3$ and LiD targets were
periodically reversed to cancel out possible false asymmetries. 

Charged particles leaving the target at laboratory angles of approximately
2.75 or 5.5 degrees were detected in  two independent
magnetic spectrometers, each with a momentum acceptance of 10 to 40 GeV.
The detector systems were similar in the two spectrometers, consisting
of two highly segmented planes of plastic scintillator hodoscopes for
tracking, and an array of 200 lead glass blocks used for additional tracking
information as well as energy measurements. Electron showers were fully
contained in these 24 r.l. blocks, while hadrons typically deposited 
one third of their energy in the several interaction lengths of the lead
glass. The ratio of lead glass energy $E$ to particle momentum $P$ was
found to be a useful quantity in distinguishing electrons from hadrons,
as illustrated in Fig. 1. Almost all electrons are characterized by
$E/P>0.8$, while about 80\% of the  hadrons are characterized by $E/P<0.6$.
The peak near $E/P=0.07$ is from muons and non-showering hadrons, both 
depositing about 0.7 GeV in the lead glass array. Also useful for particle
identification were the two several-meter-long gas threshold Cherenkov
counters in each spectrometer. The nitrogen gas pressure in each tank 
was set for thresholds of 69 (58) MeV for electrons, 
corresponding to 14 (12) GeV for muons, 
19 (16) GeV for pions, 57 (48) GeV for kaons, and 133 (112) GeV for protons
in the 2.75 (5.5) degree spectrometer.  These thresholds were chosen to
correspond to the point where the pion to electron 
ratio drops below unity (see Fig. 2).
The efficiency of the Cherenkov counters
was approximately 95\% for electrons, while 
the probability of a hadron below
Cherenkov threshold to produce one or more photoelectrons 
 was reduced to less than 1\% by the addition of 10\%
methane to quench scintillation light. 

The readout of all the detectors was done once per beam pulse (120 Hz), 
with hodoscope and lead glass hits registered in multi-hit TDCs, the lead
glass energies  recorded in ADCs, and the time distribution of each Cherenkov
counter response digitized in flash-ADCs at 1 nsec time intervals. Unlike
other SLAC experiments which used specific triggers to selectively record 
electron events, the present  system allowed most of the much more
copious hadron tracks and shower clusters to be reconstructed in addition
to the electron candidates. 

 Fig. 2 shows the calculated ratio of positrons, muons,
pions, kaons, and (anti-)protons rates to electron rates, 
for both the negative and positive 
polarity spectrometer settings (approximately 20\% of the data were taken
with the positive settings). The hadron rates are from a fit to
previous lower energy data \cite{wiser}. These 
predictions are in good agreement with the PYTHIA Monte Carlo \cite{pythia},
except for the $K^+$ rate, which is lower in PYTHIA by a factor of two. 
The rates are only for pions and kaons that do not decay in the
40 (25) m active length of the 2.75 (5.5) degree spectrometer. 
The electron rates are from a standard fit to deep-inelastic electron
scattering data \cite{NMC}, and include radiative tail effects. The positron  
rates are from the PYTHIA Monte Carlo for decay sources (the
most important being $\pi^0$ and $J/\psi$ decay), and a Bethe-Heitler
code \cite{bh} for the pair production contribution. The muons come
primarily from decay of pions and kaons in flight, with additional contributions
from Bethe-Heitler and $J/\psi$ decay at high transverse momenta. 

Although the detector systems were very good at identifying electrons in the
presence of a large hadron flux, the 
identification of different hadrons is not
particularly good. We therefore developed one set of cuts to define inclusive
hadrons, and a more restricted set of cuts to identify a sample that is
almost entirely pions. 
The set of cuts we used to identify inclusive
hadrons was $E>1.5$ GeV (to eliminate muons), $E/P<0.6$ (to remove electrons
and positrons), $P<29$ (24) GeV at 2.75$^\circ$ (5.5$^\circ$), 
and that the lead glass cluster not be
near the edge of the array (to avoid leakage out the sides). 
For this ``hadron'' definition, the Cherenkov information was ignored. 
The upper momentum cut was found to be necessary because the hadron
cross section drops very rapidly with increasing momentum, and contributions
from a variety of background processes begin to dominate above the momentum cut
used. Using the rates shown in Fig. 2, the remaining ``hadron'' sample
then consists of roughly 95\% pions for the negative sample, and about 70\%
for the positive sample, with the remainder about equally divided among
protons and kaons. The ``pion'' definition
required both Cherenkov counters to have a signal,
and the momentum to be above 19 (16) GeV (pion Cherenkov threshold),
  in addition to the cuts used for the ``hadron'' definition. 

Within the uncertainties of the detector efficiencies, the ratio of total 
hadron to electron rates was observed to be consistent with the 
predictions shown in Fig. 2. The ratio of positive to negative hadrons
in the momentum range 10 to 20 GeV was observed to be approximately
constant at 1.3 in both  spectrometers and for both targets, in 
rough agreement with the predictions of Fig. 2. 

The helicity-dependent asymmetries were determined according to:
\begin{equation}
A_\| =\bigg({N_- -N_+ \over
N_- +N_+} \bigg) {1 \over f' P_b P_t },
\end{equation}
where positive target polarization is defined to be parallel
to the electron beam direction, $N_-$ 
($N_+$) is number of detected hadrons  per incident charge for negative
(positive) beam helicity, and $f'$ is the dilution factor 
representing the fraction of
measured events originating from polarizable protons or 
deuterons within the
target (including the effective deuteron in $^6$Li). 
The asymmetry $A_\|$ has previously  been designated $E$ \cite{barker}
in the literature.
In calculating $f'$, we assumed the yield of hadrons per nucleon to
be independent of atomic number $A$. Possible shadowing corrections
were looked for by parameterizing the cross section per nucleon as
$A^\alpha$. Using data taken periodically during 
the experiment with targets of 
carbon, beryllium, empty cup (mostly aluminum), and empty cup filled with
helium, we found $\alpha=0.0\pm0.1$ for both spectrometers in the 
momentum range $10<P<20$, where the hadron rates are highest.
We used values of $f'=0.13\pm0.03$ for the NH$_3$ target and
$f'=0.34\pm0.04$ for the LiD target. The large error bars on $f'$ are
dominated by the large uncertainty in the nuclear dependence 
of the hadron yields, compared to the much better known $A$-dependence
for deep inelastic electron scattering. 

The results for $A_\|$ are displayed in Fig. 3 for the proton target,
and Fig. 4 for the deuteron target, for both positive and negative signed
hadrons in each spectrometer. Both the inclusive hadron and identified
pion definitions are shown. The errors shown are statistical
only. The relative systematic errors are approximately 20\% (12\%) for
the proton (deuteron) target, dominated by the uncertainty in $f'$.
The results for the proton show a significant
positive asymmetry, which is twice as large for positive hadrons as for
negative hadrons in a given spectrometer. 
The transverse momentum range in the 2.75 degree spectrometer is
$0.5<P_t<1.5$ GeV, half that of the 5.5 degree spectrometer. 
No significant difference is
seen between the inclusive hadron and identified pion results, although
the later covers a more restrictive momentum range, and has larger
statistical errors. 

 The inclusive hadron  asymmetries for the proton target are much 
smaller than the corresponding 
asymmetries for deep inelastic electrons (about 0.08 (0.16) 
for the 2.75 (5.5) degree spectrometer). Several checks were made 
(for example, by changing the $E/P$ cuts)
to ensure that the small, non-zero asymmetries are not due to contamination
of electrons or muons mis-identified as hadrons. As a check against
possible false asymmetries, it was verified that  the physics asymmetry remains
constant when the sign of the target polarization was reversed, either by
changing the direction of the magnetic field, or by changing the 
microwave frequency used in dynamic nuclear polarization. The results for
the deuteron target are all consistent with zero. 
Taken together, the deuteron and proton results imply a small negative
asymmetry for polarized neutrons. Negative asymmetries were in fact
observed for a polarized neutron target 
in both spectrometers in experiment E154 \cite{yury},
with an average value of $A_\|=-0.004\pm0.001$ for
negative hadrons, and   $A_\|=-0.008\pm0.002$ for positive hadrons,
averaged over the full momentum range of both spectrometers.  
The results  for both proton and deuteron targets are also
consistent with the results for SLAC E143 \cite{E143},
taken at lower beam energies but with considerably larger statistical errors.

Also shown in Figs. 3 and 4 are some of the calculations \cite{carlson} 
made to match the experimental conditions of 
the 5.5 degree spectrometer. The loose dotted, tight dotted, and dashed
curves all use the BBS polarized gluon distribution \cite{BBS}, with
three commonly used polarized quark distribution functions. The solid
curve is the same as the loose dotted curve, but has $\Delta G(x)=0$. 
The calculations were not done for the 2.75 degree spectrometer due to the
low average $P_t$, where Vector Meson Dominance contributions might be
expected to become important.  The
calculations take into account the energy and polarization dependence of 
the photon spectrum, but are only for pions. This should not make 
much difference for the negative signed hadrons, which are almost
all pions, as indicated in Fig. 2. The calculations are done in leading order 
perturbative QCD and take into account direct contributions, pions from
fragmentation of quarks and gluons, and resolved photon contributions. For
the range $10<P<20$ GeV where we have good measurements of $A_\|$, the
fragmentation process is dominant. Since this includes quarks produced via
the photon-gluon fusion subprocess, for which the analyzing power  is very large,
there is expected to be good sensitivity to the polarized gluon 
distribution, compared to measurements of $g_1$, where the gluon polarization
does not contribute in leading order. To compare with our deuteron data,
we have naively taken the average of the predicted asymmetries for neutron
and proton targets. 

In general, the calculations predict  larger asymmetries than
observed experimentally, especially for the positive hadrons from the
proton target. In this case, the differences between the curves are less
than the differences between any of the curves and the data. This 
makes it impossible to draw any conclusions about $\Delta G(x)$.
The calculations for the deuteron do predict smaller 
asymmetries than for the proton, as observed experimentally, but also
tend to be higher than the data for positive hadrons. 
None of the choices of quark and gluon polarization are in good agreement
with our entire data set.  It is possible that soft
processes, not easily calculable in perturbative QCD, are playing a
more significant role than expected in the calculations. 
Extending the calculations to NLO and the inclusion of soft processes
(in progress by the authors of
\cite{carlson}) may also lead to better agreement with the data. Ideally,
the calculations should also include kaons and protons in the final state.
It thus
remains as an interesting theoretical challenge to calculate the full
gamut of processes in inclusive polarized hadron photoproduction. The
present data will provide valuable experimental constraints on such models,
and perhaps lead to constraints on the gluon polarization in the nucleon
in the future. 

We thank the authors of Ref. \cite{carlson} for valuable discussions and
for performing calculations at the kinematics of this experiment.
This work was supported by the Department of Energy
(TJNAF, Massachusetts, ODU, SLAC, Stanford, Virginia, 
William and Mary, and Wisconsin);  by the National Science
Foundation (American, Kent State, Michigan and ODU); 
by the Schweizersche Nationalfonds (Basel); 
by the Commonwealth of Virginia (Virginia); 
by the Kent State University Research Council (GGP); 
by the Centre National de la Recherche Scientifique and 
the Commissariat a l'Energie Atomique (French groups).

\vfill\eject
\begin{figure}
\vspace*{6.2in}
\hspace*{.45in}
\includegraphics{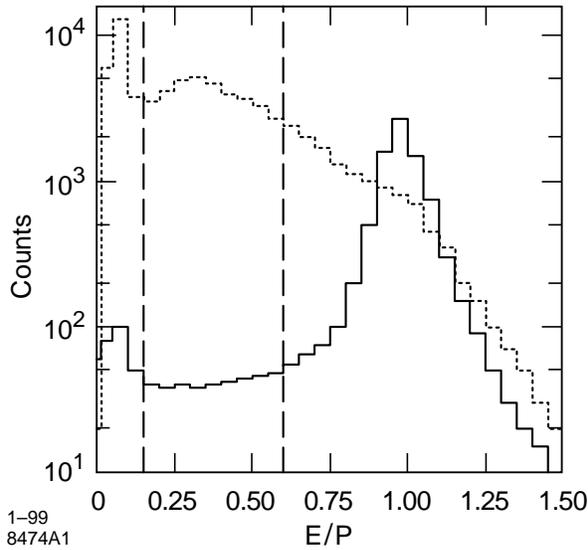}
\caption{Number of counts as a function of the ratio of 
calorimeter energy $E$ to 
track momentum $P$ for a typical 1 hour data run with the 5.5 degree
spectrometer set in negative polarity. The solid curve is for the case
where both Cherenkov counters had significant pulse height (mostly
electrons), while the dashed curve is for the case where the Cherenkov
counters had zero or small pulse height (mostly pions, kaons, and muons).
The momentum range is restricted to $10<P<15$ GeV. The vertical dashed
lines approximately show the cuts used to define the hadron sample. } 
\end{figure}

\vfill\eject
\begin{figure}
\vspace*{7.2in}
\hspace*{.45in}
\includegraphics{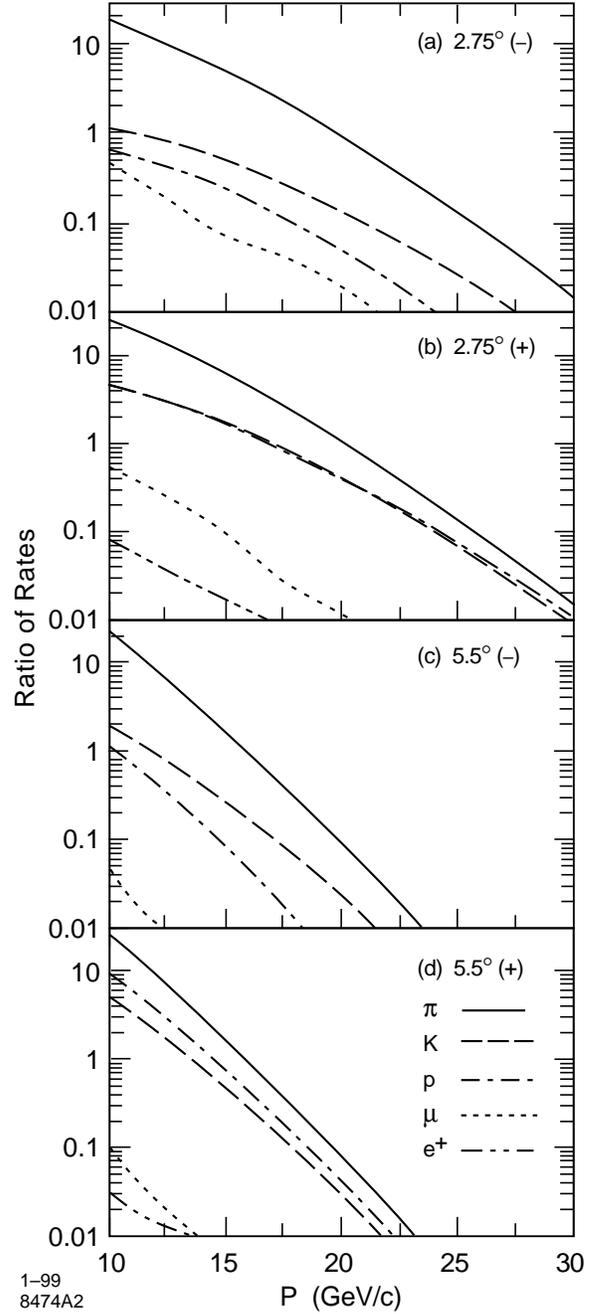}
\caption{Predicted ratio of rates of particles to the 
corresponding electron rate in each spectrometer, for positive 
and negative particles. 
The hadron rates are from Ref. \protect\cite{wiser}.}
\end{figure}

\vfill\eject
\begin{figure}
\vspace*{7.2in}
\hspace*{.45in}
\includegraphics{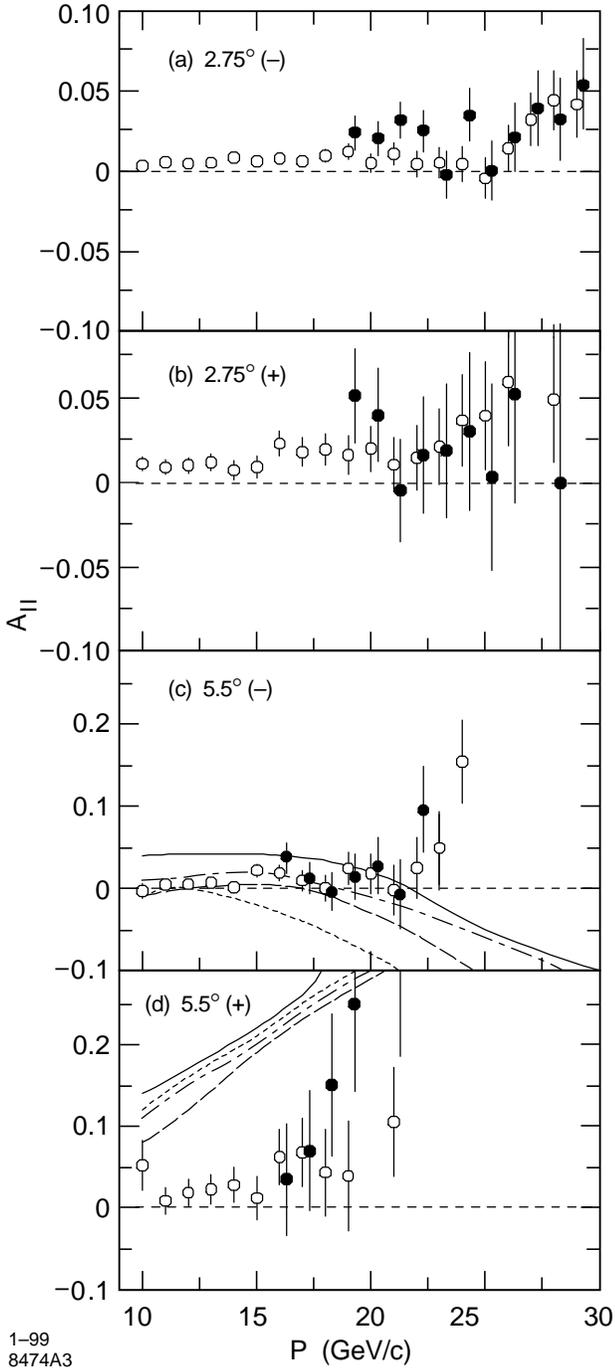}
\caption{The helicity-dependent asymmetries $A_\|$ for polarized
photoproduction of inclusive hadrons (open circles) and  
 pions (solid circles) from
a longitudinally polarized proton, for both spectrometers and for
both positive (+) and negative (-) particles. 
The 5.5 degree curves are taken from  Fig. 9 of  Ref. \protect\cite{carlson}. }
\end{figure}

\vfill\eject
\begin{figure}
\vspace*{7.2in}
\hspace*{.45in}
\includegraphics{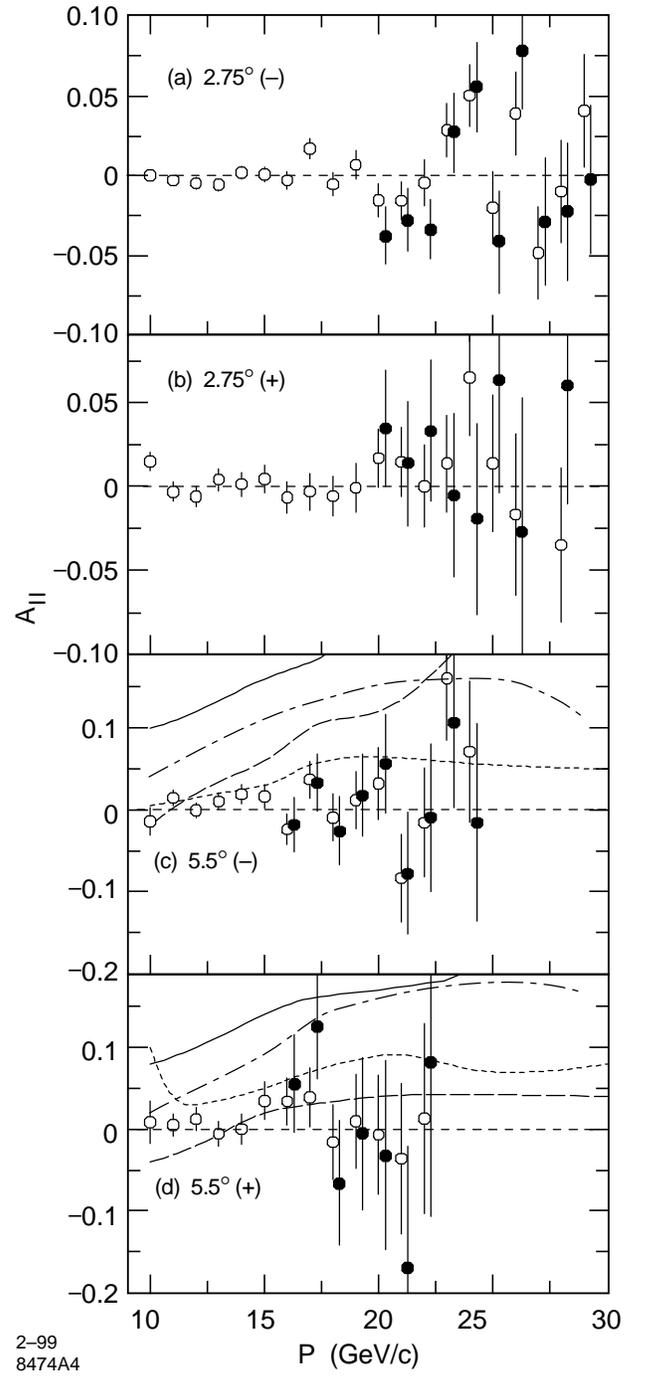}
\caption{Same as Fig.~3 except for a polarized deuteron target.
The 5.5 degree curves are an average of the neutron (Fig. 8) and proton
(Fig. 9) calculations  of Ref. \protect\cite{carlson}.}
\end{figure}
\onecolumn

\begin{table}[t]
\caption{The helicity-dependent asymmetries $A_\|$ for polarized
photoproduction of inclusive hadrons  from
a longitudinally polarized proton, for both spectrometers and for
both positive ($h^+$) and negative ($h^-$) particles. The photon endpoint
energy is 48.35 GeV. The errors are statistical only.}
\begin{tabular}{ccccc}  $P (GeV)$ & 
$\theta=2.75^{\circ}$ $h^-$ & 
$\theta=2.75^{\circ}$ $h^+$ & 
$\theta=5.5^{\circ}$ $h^-$ & 
$\theta=5.5^{\circ}$ $h^+$   \\ \hline
  10.0 & $ 0.003\pm 0.002$ & $ 0.011\pm 0.004$ & $-0.003\pm 0.009$ & $ 0.052\pm 0.031$\\
  11.0 & $ 0.006\pm 0.002$ & $ 0.009\pm 0.004$ & $ 0.005\pm 0.005$ & $ 0.009\pm 0.016$\\
  12.0 & $ 0.005\pm 0.002$ & $ 0.010\pm 0.005$ & $ 0.005\pm 0.005$ & $ 0.019\pm 0.017$\\
  13.0 & $ 0.005\pm 0.002$ & $ 0.012\pm 0.005$ & $ 0.007\pm 0.006$ & $ 0.023\pm 0.018$\\
  14.0 & $ 0.009\pm 0.002$ & $ 0.007\pm 0.006$ & $ 0.001\pm 0.006$ & $ 0.028\pm 0.022$\\
  15.0 & $ 0.006\pm 0.003$ & $ 0.009\pm 0.006$ & $ 0.022\pm 0.008$ & $ 0.012\pm 0.027$\\
  16.0 & $ 0.008\pm 0.003$ & $ 0.023\pm 0.007$ & $ 0.019\pm 0.010$ & $ 0.062\pm 0.033$\\
  17.0 & $ 0.006\pm 0.004$ & $ 0.018\pm 0.008$ & $ 0.010\pm 0.012$ & $ 0.068\pm 0.042$\\
  18.0 & $ 0.010\pm 0.004$ & $ 0.020\pm 0.009$ & $ 0.000\pm 0.016$ & $ 0.044\pm 0.053$\\
  19.0 & $ 0.012\pm 0.005$ & $ 0.017\pm 0.012$ & $ 0.024\pm 0.020$ & $ 0.039\pm 0.067$\\
  20.0 & $ 0.005\pm 0.006$ & $ 0.020\pm 0.014$ & $ 0.018\pm 0.024$ & -\\
  21.0 & $ 0.011\pm 0.007$ & $ 0.011\pm 0.016$ & $-0.002\pm 0.030$ & $ 0.105\pm 0.067$\\
  22.0 & $ 0.005\pm 0.008$ & $ 0.015\pm 0.019$ & $ 0.025\pm 0.038$ & -\\
  23.0 & $ 0.005\pm 0.009$ & $ 0.021\pm 0.023$ & $ 0.050\pm 0.044$ & -\\
  24.0 & $ 0.005\pm 0.011$ & $ 0.037\pm 0.027$ & $ 0.155\pm 0.051$ & -\\
  25.0 & $-0.004\pm 0.013$ & $ 0.040\pm 0.032$ & - & -\\
  26.0 & $ 0.014\pm 0.015$ & $ 0.060\pm 0.038$ & - & -\\
  27.0 & $ 0.033\pm 0.017$ & - & - & -\\
  28.0 & $ 0.045\pm 0.018$ & $ 0.049\pm 0.037$ & - & -\\
  29.0 & $ 0.042\pm 0.021$ & - & - & -\\
 \end{tabular}
\end{table}

\begin{table}[t]
\caption{Same as Table I, but for identified pions.}
\begin{tabular}{ccccc}  $P (GeV)$ & 
$\theta=2.75^{\circ}$ $\pi^-$ & 
$\theta=2.75^{\circ}$ $\pi^+$ & 
$\theta=5.5^{\circ}$ $\pi^-$ & 
$\theta=5.5^{\circ}$ $\pi^+$   \\ \hline
  16.0 & - & - & $ 0.037\pm 0.018$ & $ 0.035\pm 0.068$\\
  17.0 & - & - & $ 0.013\pm 0.020$ & $ 0.070\pm 0.073$\\
  18.0 & - & - & $-0.004\pm 0.023$ & $ 0.151\pm 0.087$\\
  19.0 & $ 0.024\pm 0.011$ & $ 0.052\pm 0.028$ & $ 0.014\pm 0.028$ & $ 0.249\pm 0.107$\\
  20.0 & $ 0.021\pm 0.011$ & $ 0.040\pm 0.028$ & $ 0.028\pm 0.034$ & -\\
  21.0 & $ 0.032\pm 0.011$ & $-0.005\pm 0.031$ & $-0.007\pm 0.042$ & $ 0.304\pm 0.119$\\
  22.0 & $ 0.025\pm 0.013$ & $ 0.016\pm 0.034$ & $ 0.096\pm 0.053$ & -\\
  23.0 & $-0.002\pm 0.015$ & $ 0.019\pm 0.040$ & - & -\\
  24.0 & $ 0.035\pm 0.016$ & $ 0.031\pm 0.047$ & - & -\\
  25.0 & $ 0.001\pm 0.019$ & $ 0.003\pm 0.055$ & - & -\\
  26.0 & $ 0.021\pm 0.021$ & $ 0.053\pm 0.065$ & - & -\\
  27.0 & $ 0.040\pm 0.024$ & - & - & -\\
  28.0 & $ 0.033\pm 0.026$ & $ 0.096\pm 0.065$ & - & -\\
  29.0 & $ 0.055\pm 0.028$ & - & - & -\\
 \end{tabular}
\end{table}

\begin{table}[t]
\caption{Same as Table I, but for deuteron target.}
\begin{tabular}{ccccc}  $P (GeV)$ & 
$\theta=2.75^{\circ}$ $h^-$ & 
$\theta=2.75^{\circ}$ $h^+$ & 
$\theta=5.5^{\circ}$ $h^-$ & 
$\theta=5.5^{\circ}$ $h^+$   \\ \hline
  10.0 & $ 0.000\pm 0.003$ & $ 0.015\pm 0.006$ & $-0.014\pm 0.017$ & $ 0.009\pm 0.026$\\
  11.0 & $-0.003\pm 0.003$ & $-0.003\pm 0.006$ & $ 0.015\pm 0.009$ & $ 0.005\pm 0.014$\\
  12.0 & $-0.005\pm 0.003$ & $-0.006\pm 0.006$ & $-0.001\pm 0.009$ & $ 0.012\pm 0.014$\\
  13.0 & $-0.006\pm 0.004$ & $ 0.004\pm 0.007$ & $ 0.010\pm 0.010$ & $-0.005\pm 0.016$\\
  14.0 & $ 0.002\pm 0.004$ & $ 0.001\pm 0.007$ & $ 0.019\pm 0.012$ & $ 0.000\pm 0.018$\\
  15.0 & $ 0.001\pm 0.005$ & $ 0.004\pm 0.008$ & $ 0.016\pm 0.015$ & $ 0.035\pm 0.023$\\
  16.0 & $-0.003\pm 0.005$ & $-0.007\pm 0.009$ & $-0.024\pm 0.018$ & $ 0.034\pm 0.029$\\
  17.0 & $ 0.017\pm 0.006$ & $-0.003\pm 0.011$ & $ 0.037\pm 0.023$ & $ 0.039\pm 0.036$\\
  18.0 & $-0.006\pm 0.007$ & $-0.006\pm 0.012$ & $-0.009\pm 0.029$ & $-0.016\pm 0.046$\\
  19.0 & $ 0.007\pm 0.009$ & $-0.001\pm 0.015$ & $ 0.012\pm 0.035$ & $ 0.009\pm 0.058$\\
  20.0 & $-0.016\pm 0.010$ & $ 0.017\pm 0.017$ & $ 0.032\pm 0.044$ & $-0.007\pm 0.072$\\
  21.0 & $-0.016\pm 0.012$ & $ 0.015\pm 0.020$ & $-0.083\pm 0.054$ & $-0.036\pm 0.092$\\
  22.0 & $-0.005\pm 0.014$ & $ 0.000\pm 0.024$ & $-0.015\pm 0.066$ & $ 0.013\pm 0.116$\\
  23.0 & $ 0.028\pm 0.017$ & $ 0.014\pm 0.029$ & $ 0.161\pm 0.076$ & -\\
  24.0 & $ 0.050\pm 0.019$ & $ 0.065\pm 0.034$ & $ 0.071\pm 0.086$ & -\\
  25.0 & $-0.020\pm 0.022$ & $ 0.014\pm 0.041$ & - & -\\
  26.0 & $ 0.039\pm 0.026$ & $-0.017\pm 0.048$ & - & -\\
  27.0 & $-0.049\pm 0.029$ & - & - & -\\
  28.0 & $-0.010\pm 0.032$ & $-0.035\pm 0.046$ & - & -\\
  29.0 & $ 0.041\pm 0.035$ & - & - & -\\
 \end{tabular}
\end{table}

\begin{table}[t]
\caption{Same as Table III, but for identified pions.}
\begin{tabular}{ccccc}  $P (GeV)$ & 
$\theta=2.75^{\circ}$ $\pi^-$ & 
$\theta=2.75^{\circ}$ $\pi^+$ & 
$\theta=5.5^{\circ}$ $\pi^-$ & 
$\theta=5.5^{\circ}$ $\pi^+$   \\ \hline
  16.0 & - & - & $-0.018\pm 0.033$ & $ 0.056\pm 0.060$\\
  17.0 & - & - & $ 0.033\pm 0.035$ & $ 0.126\pm 0.064$\\
  18.0 & - & - & $-0.026\pm 0.042$ & $-0.065\pm 0.076$\\
  19.0 & - & - & $ 0.018\pm 0.050$ & $-0.005\pm 0.093$\\
  20.0 & $-0.038\pm 0.018$ & $ 0.035\pm 0.034$ & $ 0.056\pm 0.061$ & $-0.032\pm 0.116$\\
  21.0 & $-0.028\pm 0.020$ & $ 0.014\pm 0.037$ & $-0.077\pm 0.074$ & $-0.169\pm 0.149$\\
  22.0 & $-0.034\pm 0.022$ & $ 0.033\pm 0.042$ & $-0.010\pm 0.090$ & $ 0.082\pm 0.188$\\
  23.0 & $ 0.026\pm 0.025$ & $-0.005\pm 0.049$ & $ 0.107\pm 0.104$ & -\\
  24.0 & $ 0.055\pm 0.028$ & $-0.020\pm 0.057$ & $-0.016\pm 0.121$ & -\\
  25.0 & $-0.042\pm 0.032$ & $ 0.063\pm 0.067$ & - & -\\
  26.0 & $ 0.078\pm 0.036$ & $-0.027\pm 0.080$ & - & -\\
  27.0 & $-0.029\pm 0.040$ & - & - & -\\
  28.0 & $-0.023\pm 0.043$ & $ 0.060\pm 0.071$ & - & -\\
  29.0 & $-0.003\pm 0.047$ & - & - & -\\
 \end{tabular}
\end{table}


\begin{references}
 
\bibitem[\diamond]{W&M} Present address: College of William and Mary, 
Williamsburg, VA 23187

\bibitem[\dag]{Grenoble}
Permanent Address: Institut des Sciences Nucl\'eaires, IN2P3/CNRS,
38026 Grenoble Cedex, France

\bibitem[\ddag]{LANL}
Present Address: Duke University, TUNL, Durham, NC 27708

\bibitem[\Box]{LANL}
Present Address: Saint Norbert College, DePeke, WI 54115

\bibitem[\flat]{DESY}
Present Address: DESY, D-22603, Hamburg, Germany

\bibitem[\S]{LLNL}
Present Address: Lawrence Livermore National Laboratory, Livermore, CA 94551
 
\bibitem{E155}
SLAC E155, P.~L.~Anthony {\it et al.,} SLAC--PUB--7983 (hep--ex/9901006); 
 SLAC-PUB-7994; SLAC--PUB--8041. 
 
\bibitem{SMC}
SMC, D. Adeva {\it et al.,} Phys. Rev. D58 (1998) 112001. 

\bibitem{hermes}
HERMES, A. Airapetian {\it et al.,} Phys. Lett. B442 (1998) 484. 

\bibitem{carlson}
A. Afanasev, C.~E.~Carlson, C.~Wahlquist,
Phys. Rev. D 58 (1998) 054007. 

\bibitem{many} D. De Florian and W. Vogelsang,
               Phys. Rev. D 57  (1998) 4376;
               B. A. Kniehl, hep-ph/9709261; 
               M. Stratmann and W. Vogelsang, hep-ph/9708243.

\bibitem{olson}
H.~Olsen and L.~C. Maximon, Phys. Rev. 110 (1958) 589.

\bibitem{LiD}
S.~L.~B${\ddot {\rm u}}$ltmann {\it et al.,} SLAC-PUB-7904 (1998); 
O.~A.~Rondon, Report No. aps1998dec15\_002 (1998), submitted to Phys. Rev. C.
 
\bibitem{wiser}
D. Wiser, PhD thesis, University of Wisconsin, 1977 
(unpublished). Fit available from S.~Rock (ser@slac.stanford.edu).

 \bibitem{pythia}
T. Sjostrand, Computer Physics Comm. { 82} (1994) 74.

\bibitem{NMC} NMC,  P. Arneodo {\it et al.,} Phys. Lett. {B364} (1995) 107.
 
\bibitem{bh}
Y. S. Tsai, Rev. Mod. Phys. { 46} (1974) 815; { 49} (1977) 421 (E);
R.~Yoshida, Nucl. Instrum. Meth. { A302} (1991) 63.

\bibitem{barker}
I.~S.~Barker, A.~Donnachie, and S.~K.~Storrow, Nucl. Phys. B95 (1975)
347. 

\bibitem{yury}
 F.~Sabatie, PhD Thesis, DAPNIA/SPhN-9803T (1998);
Yu.~G.~Kolomensky, PhD thesis, University of Massachusetts (1997),
(unpublished).

\bibitem{E143}
SLAC E143, K.~Abe {\it et al.,} Phys. Rev. D58 (1998) 112003.

\bibitem{BBS}
S. J. Brodsky, M. Burkardt, and I. Schmidt, Nucl. Phys. {B441} (1995), 197.

\end{references}
\end{document}